\documentclass[sigconf]{acmart}

\AtBeginDocument{%
  }

\newcommand{\tool}{SPFinder }
\usepackage{multicol,multirow}
\usepackage[most]{tcolorbox}
\usepackage{subfigure}
\usepackage{float}
\begin{document}

\settopmatter{
  printacmref=false,
  printccs=false,
  printfolios=true
}   
\renewcommand\footnotetextcopyrightpermission[1]{} 
\setcopyright{none}

\title{SPFinder: Improving the Context Length and Scalability for Tracing Known Vulnerability Patches}

\author{Jiangrui Zheng}
\email{jzheng36@stevens.edu}
\affiliation{%
  \institution{Stevens Institute of Technology}
  \city{Hoboken}
  \state{NJ}
  \country{USA}
}

\author{Xueqing Liu}
\email{xliu127@stevens.edu}
\affiliation{%
  \institution{Stevens Institute of Technology}
  \city{Hoboken}
  \state{NJ}
  \country{USA}
}

\author{Guanqun Yang}
\email{gyang16@stevens.edu}
\affiliation{%
  \institution{Stevens Institute of Technology}
  \city{Hoboken}
  \state{NJ}
  \country{USA}
}

\author{Siyan Wen}
\email{swen4@stevens.edu}
\affiliation{%
  \institution{Stevens Institute of Technology}
  \city{Hoboken}
  \state{NJ}
  \country{USA}
}

\author{Qiushi Liu}
\email{qiushi3@illinois.edu}
\affiliation{%
  \institution{ZJU-UIUC}
  \city{Haining}
  \state{Zhejiang}
  \country{China}
}

\author{Xiaoyin Wang}
\email{xiaoyin.wang@utsa.edu}
\affiliation{%
  \institution{University of Texas, San Antonio}
  \city{San Antonio}
  \state{Texas}
  \country{USA}
}



\begin{abstract}

An upstream task for vulnerability management is the accurate localization of the patch that fixes a vulnerability. Existing works have proposed several approaches to trace or retrieve the patching commit for fixing a CVE~\cite{tan2021locating,wang2022vcmatch,dunlap2024vfcfinder,li2024patchfinder,ran2025efficient}. However, they suffer from two major challenges: (1) They cannot effectively handle the long diff code in patch commits, which is common when commit messages are non-informative; and (2) they do not scale to the full repository with satisfactory accuracy in realistic settings.

We propose \tool, a scalable and effective retrieval framework for tracing known vulnerability patches. To address the long-context challenge, \tool introduces a hierarchical embedding technique that efficiently extends context coverage while mitigating long-context degradation, enabling effective modeling of all files in the commit. To address the scalability challenge, \tool adopts a three-phase retrieval framework that balances effectiveness and efficiency, achieving high recall at the full-repository level.

Our evaluation on two datasets shows that \tool outperforms state-of-the-art patch tracing methods (PatchFinder~\cite{li2024patchfinder}, PatchScout~\cite{tan2021locating}, VFCFinder~\cite{dunlap2024vfcfinder}) by a large margin, and surpasses VoyageAI~\cite{voyage}, the top commercial code embedding model, on MRR and Recall@10 by 18\% and 28\% respectively. Using \tool, we have successfully traced and merged patch links for 35 CVEs in the GitHub Advisory Database~\cite{githubAD}, demonstrating its practical applicability. An ablation study further confirms that hierarchical embedding is a practically effective solution for handling long context in patch retrieval. Our artifacts and online demo are publicly available at 
\url{https://github.com/AnonySE26/SPFinder} and \url{http://spfinder.org/}.

\end{abstract}

\maketitle

\thispagestyle{plain}
\pagestyle{plain}
\fancyhead{}
\fancyfoot{}

\section{Introduction}

\begin{figure*}[t]
    \centering
           \centering
           \vspace{-6pt}
           \includegraphics[width=0.8\linewidth,trim=0 00 0 0,clip]{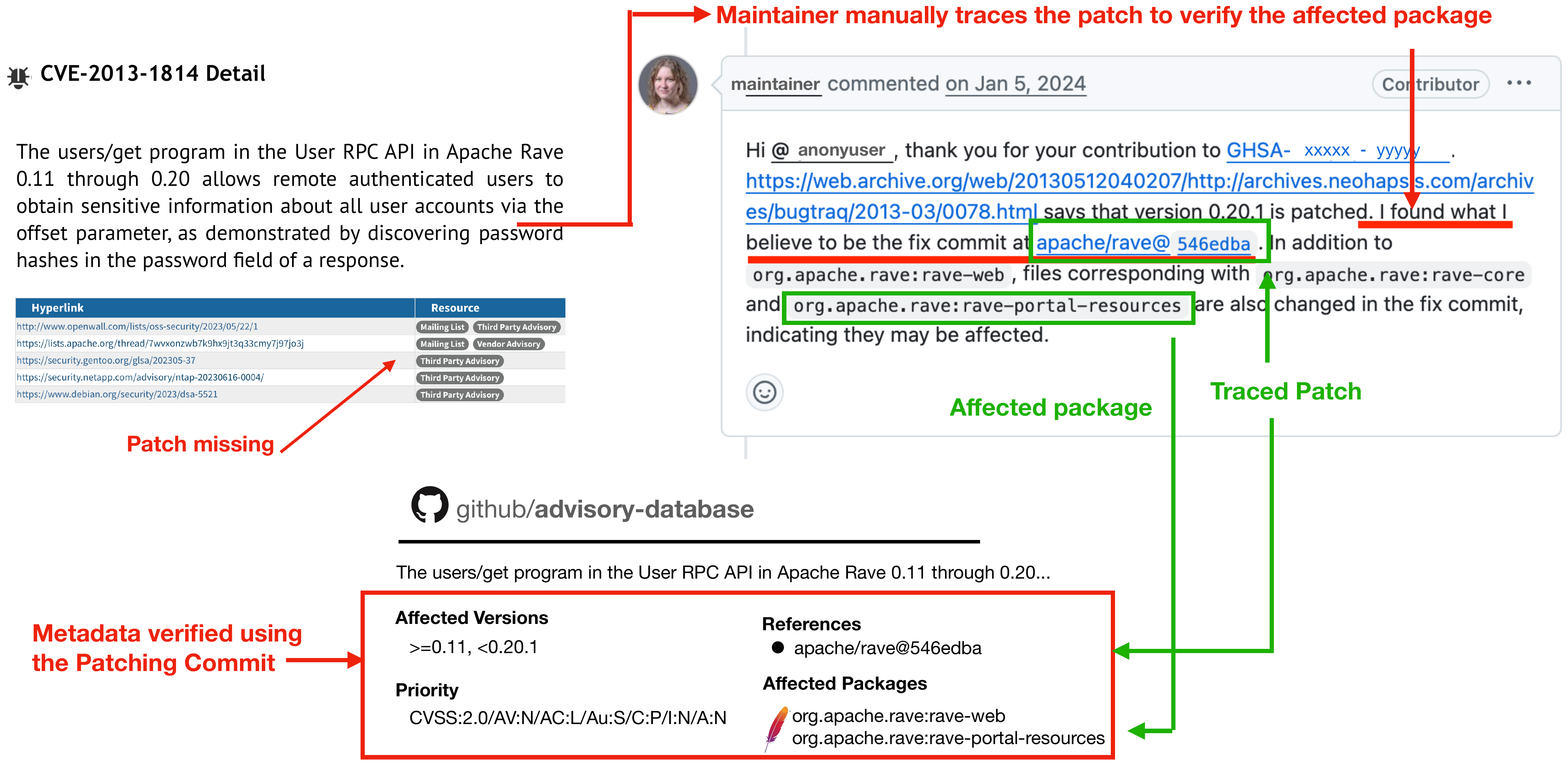}
            \vspace{-4pt}      
          \caption{Motivating example of a missing patch link for CVE-2013-1814: the NVD lacks a patch reference, a maintainer manually searches the \texttt{apache/rave} repository, and the patch is later verified in the GitHub Advisory Database. \label{fig:cve_example}}
       
   \end{figure*}

With the increasing risk of security vulnerabilities, software developers rely on vulnerability advisory databases to mitigate potential risks. A critical upstream task for vulnerability management is to localize the commit that fixes a vulnerability. Precisely localizing the patching commit is essential for analyzing
critical security information, including the affected version range~\cite{bao2022v}, software artifacts (Figure~\ref{fig:cve_example})\cite{chen2023identifying,chen2023vullibgen,chronos,lightxml}, and severity assessment score\cite{le2021deepcva}. Conversely, a missing or delayed patch link increases the difficulty of reachability analysis~\cite{sun2024tracing,zhang2024vulnerability}, and raises the risk of delayed mitigation and incorrect vulnerability assessment. 
Figure~\ref{fig:cve_example} illustrates a real-world case where a missing patch link
caused a 10-year delay in identifying affected packages.

A previous study~\cite{dunlap2024vfcfinder} found that the patch link is missing in more than 63\% of CVEs in the GitHub Advisory Database and NVD~\cite{nvd,nvd_delay,nvd_delay2,nvd_delay3}, with a significant proportion of these links being traceable—meaning the patching commit exists but is not linked.
To recover missing links, existing works propose various approaches to retrieve the patching commit given the CVE description~\cite{sun2024tracing,li2024patchfinder,tan2021locating,shen2023patchmatch,wang2022vcmatch,dunlap2024vfcfinder,zhang2024dual,xu2022tracking,ran2025efficient}. For example, PatchScout~\cite{tan2021locating} leverages 22 features and a learning-to-rank framework to rank the commits in a repository; PatchFinder~\cite{li2024patchfinder} employs a two-phase re-ranking framework combining TF-IDF and a CodeReviewer language model~\cite{li2022codereviewer,zhang2019bertscore}; and VFCFinder~\cite{dunlap2024vfcfinder} improves scalability by filtering commits between the fixed version and the previous one.

Despite these efforts, two limitations remain in achieving both efficiency and effectiveness in patch tracing.
(1) Existing methods cannot effectively handle the long diff code of a commit, which is critical when the commit message is non-informative. Some works (e.g., PatchFinder) have used language models on the code diff; however, they rely on older models with a 512-token context window (e.g., BERT~\cite{wang2022vcmatch}, CodeBERT~\cite{dunlap2024vfcfinder}, BertScore + CodeReviewer~\cite{li2024patchfinder}). Given that an average diff contains 15,000 tokens, these methods must discard a large amount of the diff tokens. While modern LLMs for embedding\footnote{We distinguish between general LLMs (e.g., CodeLlama, StarCoder, ChatGPT) and LLMs for embedding (e.g., VoyageAI, GritLM, CodeXEmbed), as the latter require specialized training; general LLMs cannot simply be used for embedding.} (e.g., VoyageAI~\cite{voyage}, CodeXEmbed~\cite{liu2024codexembed}, OpenAI code embedding~\cite{openai_embedding}) support much longer contexts, simply applying an LLM with a long context window is insufficient, due to the high computational cost of LLM inference.
(2) No existing work, to our knowledge, achieves satisfactory accuracy when scaling to the full repository. Most prior works~\cite{li2024patchfinder,tan2021locating,wang2022vcmatch} report recall@K scores using only 5,000 randomly sampled negative commits. This evaluation method significantly underestimates the runtime and overestimates retrieval accuracy on large repositories (38\% of repositories in our data have more than 5,000 commits).

To address these challenges, we propose \tool (\underline{S}calable \underline{P}atch \underline{Finder}), a scalable and effective system for tracing known vulnerability patches.
First, \textbf{to address the long-context challenge (RQ2 in Section~\ref{sec:evaluation_result})}, we introduce a \textbf{hierarchical embedding} technique. Specifically, \tool splits the code diff into files, embeds each file, and aggregates the top-K embeddings in multiple ways.
Second, \textbf{to address the scalability challenge (RQ1 in Section~\ref{sec:evaluation_result})}, we design a three-phase framework (Figure~\ref{fig:overview}):
(1) efficiently pre-rank the full repository using ElasticSearch and commit–CVE time affinity to achieve high recall;
(2) for the top-10k commits, extract 10 features (Table~\ref{tab:feature_set}) to improve efficiency; and
(3) apply learning-to-rank with hyperparameter tuning~\cite{wang2021flaml} on these features.
In Step (2), \tool also proposes a \textbf{path embedding} feature to bridge mismatches between the CVE and patch by incorporating repository-level path information.

We evaluate \tool on two datasets: (1) PatchFinder~\cite{li2024patchfinder}, and (2) GitHubAD, a dataset we built from the GitHub Advisory Database. Results (Table~\ref{tab:main_result}) show that \tool achieves recall@10 scores of 0.736 and 0.577, respectively, significantly outperforming existing patch retrieval algorithms~\cite{li2024patchfinder,tan2021locating,dunlap2024vfcfinder}. Moreover, \tool outperforms VoyageAI~\cite{voyage}, the top commercial code retrieval model, by 18\% and 28\% on MRR and Recall@10, respectively. \tool takes approximately 84 seconds to query every 10,000 commits, making it practical for deployment. An ablation study shows that hierarchical embedding with a smaller context window (e.g., 512 tokens) is an effective solution for handling long context. Finally, using \tool, we have successfully traced and merged 35 missing patch links into the GitHub Advisory Database.

The contributions of \tool are as follows:
\begin{itemize}
\item First work to investigate long-context patch tracing.
\item First open-source patch tracing tool to balance scalability with high recall at the full-repository level.
\item Proposes two new feature types: \textbf{hierarchical embedding} and \textbf{path embedding}. Experiments show that hierarchical embedding is an effective solution for handling long context in patch tracing.
\item Achieves state-of-the-art results, with Recall@10 surpassing existing patch tracing methods~\cite{li2024patchfinder,tan2021locating,dunlap2024vfcfinder} by a large margin, and outperforming the top commercial embedding model (VoyageAI) by 18\% and 28\% on two datasets.
\item Releases all artifacts at \url{https://github.com/AnonySE26/SPFinder}.
\end{itemize}

\section{Background}

\label{sec:existing_work_limitation}


\subsection{Known Patch Tracing: Problem Formulation}

Given the natural language description $CVE$ of a security vulnerability (e.g., Figure~\ref{fig:cve_example}) and the repository it locates in (e.g., \texttt{apache/rave} Figure~\ref{fig:cve_example}), the problem that \tool tries to solve is how to \emph{retrieve} the commit(s) $commit$ which fix(es) the vulnerability $CVE$, among all the commits in the repository. 

\noindent \textbf{Difference with Silent Fix Identification}. Based on the previous peer-review feedback, we hope to respectfully point out that our problem is \emph{not} to be confused with vulnerability detection or \emph{silent patch identification}, e.g., VulFixMiner~\cite{Zhou2021FindingAN}, RepoSPD~\cite{Wen2024RepositoryLevelGR}. The latter problem is a binary classification problem and it does not leverage the CVE. Therefore, there is no need to compare \tool with these methods (Section~\ref{sec:evaluation_result}). 

\noindent \textbf{Existing Work}
There already exist several works on known vulnerability patch tracing~\cite{wang2022vcmatch,dunlap2024vfcfinder,li2024patchfinder,tan2021locating,ran2025efficient}. 
Representative approaches include PatchScout~\cite{tan2021locating},
PatchFinder~\cite{li2024patchfinder}, and VFCFinder~\cite{dunlap2024vfcfinder}. More details about these methods are provided in Appendix~\ref{app:baseline}.

\noindent \textbf{Broader Context.}
Patch tracing is an important upstream task for vulnerability metadata curation
in security advisory databases, such as NVD and the GitHub Advisory Database~\cite{nvd,githubAD},
which are widely used for managing supply-chain risks~\cite{log4shell}.
The patching commit provides critical evidence for identifying affected software
and packages~\cite{yang2021few,dong2022survey,anwar2021cleaning,kuehn2021ovana,
lightxml,fastxml,chronos,chen2023identifying,chen2023vullibgen,wu2024identifying},
narrowing affected version ranges via version-history analysis~\cite{bao2022v,wu2024vision},
and assessing vulnerability severity from code changes~\cite{le2021deepcva}.
Beyond repository-level retrieval, prior work has also studied patch tracing via
reference-link crawling~\cite{xu2022tracking} and file-level localization~\cite{sun2024tracing,zhang2024vulnerability}.
Meanwhile, recent advances in embedding-based retrieval enable scalable similarity
search between CVE descriptions and commits using dense representations and
vector indexing~\cite{xiong2020approximate,karpukhin2020dense,khattab2020colbert,malkov2018efficient},
with state-of-the-art embedding models provided by both commercial~\cite{voyage}
and open-source systems~\cite{muennighoff2022mteb}.

\subsection{Open Challenges: Scalability and Long Context}

Despite the existing work, there remain the following challenges: 

\noindent \textbf{Challenge 1: Existing Open-Source Framework Cannot Scale to the Full Repository with Satisfactory Accuracy}.  
Another limitation of existing approaches is that, for larger repositories, many methods (e.g., PatchFinder~\cite{li2024patchfinder}, PatchScout~\cite{tan2021locating}) evaluate retrieval by randomly sampling 5,000 negative examples, rather than searching the full repository. This approach underestimates the time cost and overestimates the result accuracy. From Table~\ref{tab:repo_cve_stats} of Appendix~\ref{app:data_stat}, 49\%-84\% of CVE queries in our two datasets (Section~\ref{sec:eval_context_length}) require retrieving patches from repositories with more than 5k commits.

VFCFinder~\cite{dunlap2024vfcfinder} addresses the scalability challenge by restricting the candidates to the commits between the fixed version and the prior version, but this strategy relies on accurate fixed-version tags. Our empirical analysis (Appendix~\ref{app:version}) shows that only 60.24\% of patching commits are covered by fixed-version tags, leaving many patches hard to retrieve accurately from the full repository.

\noindent \textbf{Challenge 2: Existing Work Cannot Effectively Handle Long Code}. To the best of our knowledge, all existing works leverage language models whose input lengths are only 512 tokens (e.g., BERT~\cite{wang2022vcmatch}, CodeBERT~\cite{dunlap2024vfcfinder}, BertScore + CodeReviewer~\cite{li2024patchfinder}), therefore they cannot effectively handle the long diff code (i.e., \textbf{non-patch}) which usually includes 12,000-19,000 tokens (Appendix~\ref{app:data_stat} Table~\ref{tab:stat}), especially when the commit message is non-informative (e.g., the commit message for CVE-2021-41079~\cite{CVE-2021-41079}'s patch is: "\emph{Improving robustness}"). We further report the statistics on the \#tokens in each \textbf{patching commit} in the two datasets of this paper in Table~\ref{tab:patch_stats} (this is for the patch, different from Table V for the average commit). This table shows that \textbf{long code is prevalent in vulnerability patches}. 

The details of these datasets are in Section~\ref{sec:evaluation}.

Modern LLMs for code embedding can handle longer contexts. Nevertheless, the challenge is not automatically solved by replacing the older models with newer LLMs in existing works for patch tracing~\cite{li2024patchfinder,tan2021locating,dunlap2024vfcfinder}. The practicality of actually using longer context is hindered by the inference speed of LLMs, which increases quadratically in theory and linearly in our experiment with the context length. It remains unclear what the optimal way is (feature engineering, model selection, context length selection, etc.) to efficiently and effectively embed a commit while preserving more contextual information. For example, \textbf{we cannot directly apply Voyage or GritLM to PatchFinder} because PatchFinder uses a BertScore framework (https://github.com/MarkLee131/PatchFinder). BertScore is an older architecture that does not handle GPU memory efficiently. When using PatchFinder with LLM like GritLM/SFR, the GPU memory will blow up.

\subsection{Open Question: How to Choose the Context Length to Balance Efficiency and Accuracy?}
\label{sec:open_question}


\noindent \textbf{Long-Context Extension for Patch Tracing is Non-Trivial}. While there exist long-context extension methods~\cite{zhu2024longembed}, they cannot be easily applied to our problem. Furthermore, our empirical study in Appendix~\ref{app:context_length} on the retrieval recall of the \textsf{apache/tomcat} repo shows that \textbf{when using more tokens to embed each commit, the patch retrieval score is surprisingly worse}. This is caused by the long-context degradation of LLM~\cite{long_context_degrade}. That is, when the context is too long, the embedding vector cannot capture all the information effectively. Furthermore, the runtime grows quadratically with the context length. As a result, \textbf{patch tracing is not trivially solved by directly applying LLM with a long context window}. It remains unclear how to balance efficiency and accuracy.

\section{Approach}

\subsection{Scalability Challenge: A 3-Phase Framework for Scalable Patch Tracing}
\label{sec:3_phase}

\begin{figure*}[t]
\centering
    \includegraphics[width = \linewidth]{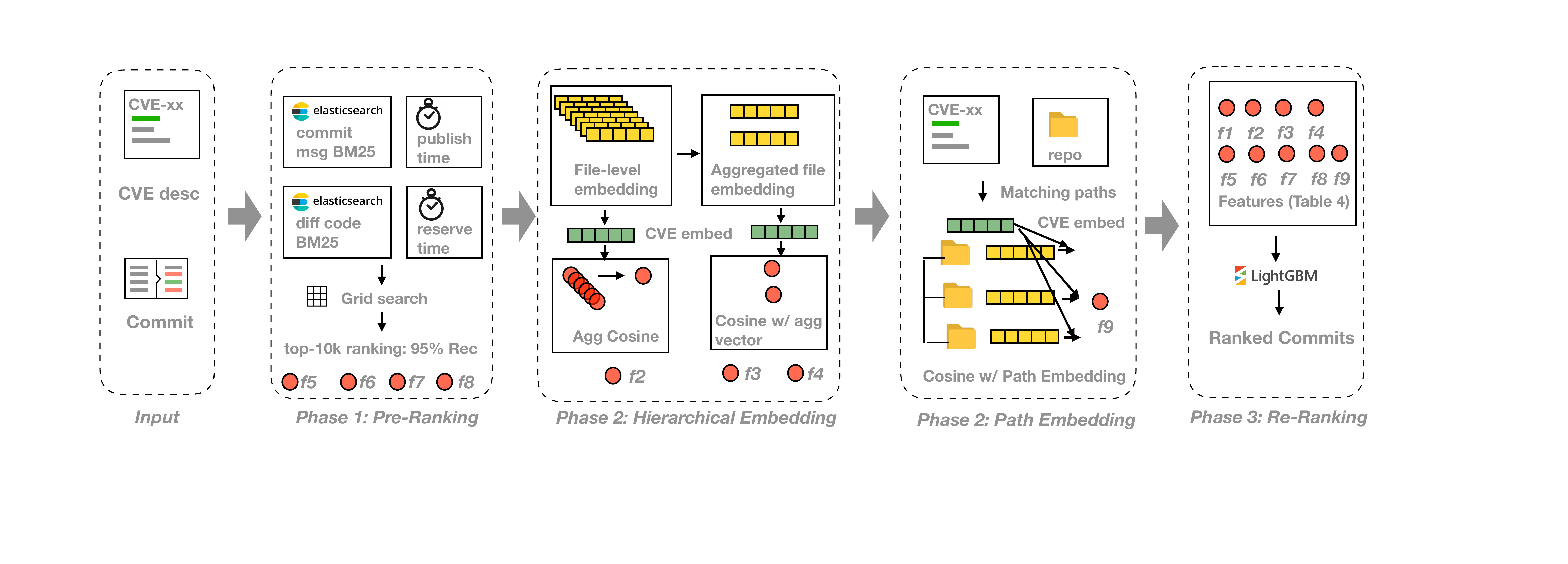}
    \caption{Overview of \tool. Given the a CVE description and all commits (commit message + code diff) of a repository, it first pre-ranks all commits using BM25 + CVE time information; then, for the top 10k commits, it extracts 10 features(Table~\ref{tab:feature_set}) including hierarchical embedding and path embedding; finally, it leverages LightGBM to combine these features into the final ranking score.\label{fig:overview}}
    \end{figure*}

The framework of \tool is shown in Figure~\ref{fig:overview}. It contains three phases as follows.

\noindent \textbf{Phase 1: Efficiently Pre-Ranking the Full Repository}. First, \tool leverages BM25 with ElasticSearch~\cite{es} (e.g., an efficient search engine using inverted index) and the CVE publish/reserve time to efficiently pre-rank the full repository, achieving a 0.95 recall@10K. The CVE time feature measures the difference (number of commits) between the CVE publish/reserve time and the commit time, based on the intuition that vulnerability fixes typically occur shortly before or after disclosure. Previous work finds that this simple feature effectively improves the patch retrieval performance~\cite{tan2021locating}. We use the publish/reserve time in the MITRE CVE database~\cite{cve}. 
The ranking signals are normalized by reciprocal rank and combined
using a weighted sum tuned on the training set; implementation details
and hyperparameter settings are provided in Appendix~\ref{app:hyperparameters}.

\noindent \textbf{Phase 2: Preparing Features for the Top-10K Commits}. After Phase 1, for each of the top-10K commits, we prepare 10 features (5 feature groups) as Table~\ref{tab:feature_set} shows. 10K is large enough to cover 95\% of the patches in the training data. Compared to previous work, we introduce two new feature groups: (1) hierarchical embedding, for extending the context length to represent the long diff code; (2) path embedding, for bridging the mismatch between CVE description and the patch. The details of these two feature groups are introduced in Section~\ref{sec:HierarchicalEmbedding} and Section~\ref{sec:NER}, respectively.

\noindent \textbf{Phase 3: Re-Ranking with Learning-to-Rank}. After preparing features, we apply learning-to-rank with the LightGBM library~\cite{ke2017lightgbm} to learn how to combine these features. More specifically, we use the LambdaRank algorithm~\cite{burges2010ranknet}. We further perform hyperparameter tuning using the FLAML library~\cite{wang2021flaml} to automatically optimize hyperparameters such as learning rate, number of leaves, and minimum data per leaf of the LambdaRank model. The model was trained to optimize evaluation metrics including NDCG, recall, and Mean Reciprocal Rank (MRR) at retrieval positions 10, 100, and 1000.
The detailed hyperparameters are shown in Appendix~\ref{app:hyperparameters}.

\begin{table}[h]
    \caption{Features used in \tool \label{tab:feature_set}}
    \centering
    \begin{tabular}{p{2.5cm}p{5.5cm}} \hline
    Feature Group & Features\\ \hline
   Commit embedding & 1. Commit-level similarity\\ \hline
     \multirow{3}{*}{Hierarch embedding}& 2. Max of file-level similarity\\ 
    & 3. Cosine with top-1 file embedding \\ 
    & 4. Cosine with top-2 file embeddings' mean \\
    \hline
    \multirow{2}{*}{BM25} & 5. BM25 commit msg\\ 
 & 6. BM25 code\\ \hline
    \multirow{2}{*}{Time} & 7. \#commits between CVE reserve time and commit \\
    & 8. \#commits between CVE publish time and commit \\ \hline
    \multirow{2}{*}{Path embedding} & 9. Jaccard Index between NER-paths and commit-paths \\
    & 10. Voyage AI~\cite{voyage} cosine between NER-paths and commit-paths \\ \hline
    \end{tabular}
    \end{table}


\subsection{Long-Context Challenge: Extending the Context Length using Hierarchical Embedding}
\label{sec:HierarchicalEmbedding}


\noindent \textbf{Scalable Embedding using LLM}. Unlike PatchFinder which couples the CVE and the commit which consumes a 3-dimensional GPU memory (i.e., batch size, code length, vector length)~\cite{zhang2019bertscore}, we adopt the sentence transformer~\cite{reimers2019sentence} style framework to decouple the embedding of CVE and commits for improving the efficiency. \tool embeds the CVE description and commit into two vectors: $v_{CVE}$ and $v_{commit}$, and ranks the commits using the cosine similarity between the vectors. We choose to use LLMs for embedding instead of older language models, e.g., CodeBERT~\cite{liu2024codexembed}, because the former generally outperforms the latter and they supports longer contexts. For example, VoyageAI~\cite{voyage}: 32,000, SalesForce's CodeXEmbed~\cite{liu2024codexembed}: 32,768. \textbf{Notice although LLM inference is slower than CodeBERT, the query is fast} because querying only requires computing the dot product of a vector the same length (1024 or 4096) as CodeBERT, therefore their query time costs are the same. 

\noindent \textbf{Prompt Engineering}. Most LLMs for embedding require a prompt template to improve the model performance. As a result, we use the following prompt template for both the CVE and file/commit embedding. The prompt for each commit: "\emph{This is a commit (commit message + diff code) of a repository. Represent it to retrieve the patching commit for a CVE description: Commit message: [commit\_msg]; Diff code: [diff]}"; the prompt for each CVE: "\emph{Represent this CVE description to retrieve the commit (commit message + diff code) that patches this CVE: [CVE desc]}". In our preliminary study with \texttt{apache/tomcat}, we find that this prompt template can improve the recall@100 of GritLM embedding by 15\% (compared to without the template).

\noindent \textbf{Which LLM to Use to Embed the Commit?} To choose an LLM for embedding, we refer to the massive text embedding benchmark (MTEB)~\cite{muennighoff2022mteb}. In particular, we select from top models on the Code Information Retrieval (CoIR)~\cite{li2024coir} and CodeSearchNet~\cite{husain2019codesearchnet} benchmarks. Among the top models in CoIR, we select the following: \texttt{voyage-3}, a model by VoyageAI~\cite{voyage}, \texttt{gritLM/GritLM-7B}, a model by ContextualAI based on Mistral-7B~\cite{jiang2024identifying}, and \\
\noindent 
\texttt{Salesforce/SFR-Embedding-Mistral}, a model by Salesforce also based on Mistral-7B, also known as CodeXEmbed~\cite{liu2024codexembed}. The information of each model is in Table~\ref{tab:model_select} of Appendix~\ref{app:model_select}.

\noindent \textbf{Hierarchical Embedding: Aggregating Top-K File Diffs}. We propose a hierarchical embedding approach that splits each diff by files, embeds the top-512 tokens of each file diff, and aggregates the embedding vectors or cosine similarity scores. Within each file, we choose to embed the top-512 tokens since they start with the file path (e.g., \texttt{diff --git a/file.java}), thus unifying the format of the embedded data which facilitates LLM's representation. We choose 512 based on the empirical analysis in Section~\ref{sec:open_question}, the LLMs achieve the best score and the highest efficiency at 512 tokens. 

When aggregating the embeddings of each file diff, we propose to only aggregate the top-K most relevant file diffs. The number of files in each commit varies from 1 to several hundred (Appendix~\ref{app:data_stat} Table~\ref{tab:stat}), while the patching information is located only in one or a few. Averaging the similarity score across all files will result in a large variance in the denominator (i.e., the number of similarity scores to aggregate) and dilute the relevant information; on the other hand, aggregating only the top-K files allows us to unify the denominator and make the scores more comparable. The file-level similarity score with the CVE description can be computed efficiently using BM25 with ElasticSearch~\cite{es} (the time cost can be found in Table~\ref{tab:efficiency}). More specifically, we propose three features (corresponding to Table~\ref{tab:feature_set}) as follows, by varying between computing the cosine score first or averaging the vector first. $v_f$ denotes the vector embedding of file $f$.  An example is provided in Table~\ref{tab:feature_values} of Appendix~\ref{app:feature_values} to help understand the meanings of features 2-4.



\noindent \textbf{Implementation Details for Speeding up the Querying}. The hierarchical embedding approach requires a large number of querying operations between the CVE and the files. Our pipeline is implemented as follows to improve the efficiency. First, for each repo, we embed all files and store the vectors with each (commit, file) pair as the key in a shared-memory store for that repo (i.e., akin to a lightweight vector database). Then, we use Python's shared-memory object in \texttt{multiprocessing} and \texttt{numpy} (e.g., \texttt{reduceat}) to compute the file-level similarity in an efficient and parallelizable way.



\subsection{Bridging CVE-Patch Mismatch using Path Embedding}
\label{sec:NER}

\noindent \textbf{Motivating Example}. One challenge for matching the CVE description and diff code is that some \emph{identifying tokens} in CVE (e.g., function names, file names) mentioned in the CVE description are not in the diff code. For example, for CVE-2021-41079~\cite{CVE-2021-41079}: "\emph{Apache Tomcat 8.5.0 to 8.5.63, 9.0.0-M1$\cdots$ to use \textcolor{red}{\textbf{NIO+OpenSSL}} or \textcolor{red}{\textbf{NIO2+OpenSSL}}$\cdots$}" (Figure~\ref{fig:ner}), the keywords are not found in the patching commit~\cite{CVE-2021-41079-patch}. On the other hand, the keywords \emph{\textcolor{blue}{\textbf{util}/\textbf{net}}} in the patch file \textit{java/org/apache/tomcat/\textcolor{blue}{\textbf{util}/\textbf{net}}/openssl/OpenSSLEngine.java} is not found in the CVE description. However, if we search for "\textbf{\textcolor{red}{\emph{NIO2}}}", the top files are located under the \emph{\textcolor{blue}{\textbf{util}/\textbf{net}}} directory. As a result, the CVE and the patch (Figure~\ref{fig:ner}) can be linked together. In the error analysis in Section~\ref{sec:rq1}, we find \textbf{CVE-Patch mismatch consist 42\% of all errors, validating the generality of the path embedding feature. }


\noindent \textbf{Path Embedding}. To bridge the gap, we first use Named Entity Recognition (NER) to extract the identifying words (variable names/file names/function names/path/etc) in the CVE description, then use GitHub's code search API~\cite{github_code_search_api} to search file paths that match these words. We then compute the similarity between the searched paths and the paths of each commit. More specifically, we use OpenAI's \texttt{GPT-4o} API for the NER and \texttt{voyage-3}~\cite{voyage} for the path embedding. The details of the features, prompt for NER, and data labeling criteria. can be found in Table~\ref{tab:feature_set} and our anonymous repository. 


\section{Evaluation Setting}
\label{sec:evaluation}


\subsection{A New Dataset: GitHubAD}

\noindent \textbf{The PatchFinder Dataset is Skewed and Outdated}. Our evaluation focuses on two datasets: (1) PatchFinder~\cite{li2024patchfinder}, and (2) GitHubAD, a new dataset we build from the GitHub Advisory database~\cite{githubAD}. The reason we need to introduce a new dataset is that the CVEs in PatchFinder are old  (1999-2022, average 2017) and exhibit a skewed distribution, heavily concentrated on a few well-known repositories. This limits the dataset’s ability to evaluate performance on newer CVEs and the ones in less known repos. 


\noindent \textbf{Introducing the GitHubAD Dataset for Fair Evaluation}. To fairly evaluate the performance, we collect the GitHubAD dataset, which contains CVEs from May 2009 to Oct 2024, average 2020, and the CVEs are more evenly distributed. For each CVE, we collect the commit URLs in the reference links as the patching commits, e.g., the patch of CVE-2021-41079 is \texttt{34115fb}\cite{CVE-2021-41079-github}. This definition follows VFCFinder~\cite{dunlap2024vfcfinder}. Arguably, the reference commits may not be officially confirmed as the patch. However, since these commits are all relevant to the CVE, a higher ranking score indicates that the model is also effective in retrieving the correct patching commit.

\noindent \textbf{Train/Test Split: Evaluating the Large Repositories}. Following previous work~\cite{chen2023identifying,dunlap2024vfcfinder}, we split both PatchFinder and GitHubAD into training and testing sets by repositories: each repository appears in either training or testing, but not both, to reduce the data contamination \footnote{Why we do not reserve the earlier CVE in the same repo for training: the earlier data within a repo for training only works after a repo has accumulated a significant amount of CVEs. 80\% repos in GitHubAD only have 1-2 CVEs.}. We reserve the repos with larger number of commits for testing. The statistics of the datasets are shown in Table~\ref{tab:stat} of Appendix~\ref{app:data_stat}. For GitHubAD's training set, we randomly sample 477 repositories to reduce the indexing cost for training.

\noindent \textbf{Preparing the Training Data for Learning to Rank}. To prepare the training data for learning-to-rank, for each CVE for training, we randomly sample a set of 1000 commits as the negative commits, including 500 hard negative commits which are the top-500 commits ranked by BM25+Time, and 500 random negative commits. \textbf{Notice we only sample the negative commits for training, but not for testing. Our test candidates are the full repository.}


\subsection{Experiment Setup}
\label{sec:experiment_setup}

Our experiments are conducted on a university-managed high-performance computing (HPC) cluster. The cluster consists of 2 nodes, each containing 4 x NVIDIA L40S GPUs of 46G GPU Ram; and 4 nodes, each containing 4 x NVIDIA H100 GPUs of 80G GPU Ram. We run the experiments of \tool on the L40S nodes, while the experiments on PatchScout, PatchFinder, and VFCFinder are run on the L40S nodes and H100 nodes. The runtime cost experiments of all methods are re-run on the L40S nodes (Table~\ref{tab:efficiency}).

\subsection{Baselines}
\label{sec:baseline}

We compare \tool with the following baselines:

\noindent \textbf{PatchFinder Phase-1}~\cite{li2024patchfinder}: We compare \tool with the Phase-1 model of PatchFinder~\cite{li2024patchfinder}. \textbf{We do not compare with Phase-2 because}: Phase-2 reranks the top-100 of Phase-1, therefore Phase-2's recall $\leq$ Phase-1's recall@100. Table~\ref{tab:main_result} shows that PatchFinder's Phase-1 recall@100 $<$ 0.219 on our two datasets, therefore no need to compare with the Phase-2 scores. \textbf{Notice while PatchFinder shows a recall of 0.78 in its stage 2, this recall is overestimated}. We provide a detailed explanation and a concrete example in
Appendix~\ref{app:patchfinder_phase2}.  
    
\noindent \textbf{PatchScout}~\cite{tan2021locating}: We compare against PatchScout using PatchFinder’s implementation~\cite{vfcfinder_github}, as PatchScout’s original implementation is unavailable; 

\noindent \textbf{VFCFinder}~\cite{dunlap2024vfcfinder}: VFCFinder requires a prohibitively higher cost than any other method (Table~\ref{tab:efficiency}), since VFCFinder is designed to operate directly with the raw directory. While it is possible for us to refactor VFCFinder, it would require a significant change of the code and it is prone to error. As a result, we compare \tool and VFCFinder on a subset of 8 repositories from GitHubAD (234 CVEs): The 8 repositories are among the most challenging repositories in GitHubAD, the complete list can be found in our anonymous repo; \\
\noindent \textbf{Voyage-3}~\cite{voyage}: For each commit, we use \texttt{voyage-3} with the top 50,000 characters (approximately 15,000 to 20,000 tokens). We use this setting to leverage Voyage's long-context capability and to save the financial cost as explained below. 

VoyageAI (acquired by MongoDB) is a commercial company focusing on embedding models to support retrieval augmented generation. As of Nov 2024, \texttt{voyage-3} was the top-1 model on the Code Information Retrieval (CoIR)~\cite{li2024coir} benchmark. The cost of \texttt{voyage-3} is \$0.06 per 1M tokens. In our experiment, the estimated cost for indexing each 10,000 commits using \texttt{voyage-3} is \$1.8. To save the cost of running \texttt{voyage-3}, we truncate each commit diff code to the first 50,000 characters (approximately 15,000 to 20,000 tokens). After the truncation, the cost for running the entire PatchFinder test dataset is approximately \$122, and for the GitHubAD test dataset is \$220. On the other hand, \texttt{voyage-code-3} and \texttt{voyage-3-large} further outperformed \texttt{voyage-3} on the CoIR. However, their prices are 3 times higher than \texttt{voyage-3}. We do not include them in our experiment since they are not cost-effective for patch retrieval.

\noindent \textbf{Why Not Compare with VCMatch}. VCMatch~\cite{wang2022vcmatch}'s score is stronger than PatchScout and weaker than PatchFinder as reported by the PatchFinder paper~\cite{li2024patchfinder}. Although VCMatch's code is available~\cite{vcmatch_code}, there remains a significant challenge in successfully compiling and reproducing the code, even after reaching out to the authors of both VCMatch and PatchFinder.

\noindent \textbf{Why Not Compare with Ran et al.\cite{ran2025efficient}}. \cite{ran2025efficient} does not open source. See the email response from Ran (permission granted): https://github.com/AnonySE26/SPFinder/blob/master/fse\_email.png.

\subsection{Evaluation Metrics}

For evaluating the effectiveness of \tool and the baselines, we use the following metrics: 

\begin{itemize}
    \item \textbf{Mean Reciprocal Rank}, $1/rank\hspace{1mm}of\hspace{1mm}first\hspace{1mm}patch$; 
    \item \textbf{Recall@k}, which evaluates the model's completeness to include all patching commits in the top-k positions. Most existing works for patch tracing focus on reporting the recall~\cite{li2024patchfinder,tan2021locating,dunlap2024vfcfinder}
    \item \textbf{NDCG@k}. The normalized discounted cumulative gain, which is similar to mean average precision (MAP), and it is a finer-grained evaluation metric than MRR and recall. 
\end{itemize}

\noindent \textbf{Why Not Evaluate Precision}. Our two datasets have 1 (PatchFinder) and 1.3 (AD) patches. That is, most CVEs only have 1 patch, making precision an uninformative metric in this context. For the same reason, existing works usually do not evaluate the precision~\cite{li2024patchfinder}. 

\section{Experimental Results}
\label{sec:evaluation_result}

\subsection{Research Questions}

Our evaluation for \tool is guided by the following research questions:

\begin{itemize}
\item \textbf{RQ1 (Section~\ref{sec:existing_work_limitation} Scalability challenge)}: How accurate/fast is \tool for tracing patches in the full repositories?
\item \textbf{RQ2 (Section~\ref{sec:existing_work_limitation} Long context challenge)}: How effective is hierarchical embedding for handling long context for patch tracing? 

\item \textbf{RQ3}: Can \tool identify real-world missing patch links?

\item \textbf{RQ4}: How does each feature in \tool contribute to the final ranking score?

\end{itemize}

\begin{table*}[th]
    \centering
    \caption{ Comparison between \tool and baselines. The t and p (paired t-test) are between \tool(sfr) and Voyage. If \tool(grit) significantly outperforms GritLM or \tool(sfr) significantly outperforms SFR ($p<$0.025), the score is labeled with $\dagger$. The best score of each row is labeled as \textbf{bold}.
    \label{tab:main_result}}
   
    \begin{tabular}{p{1.8cm} p{1.8cm}p{0.85cm} p{0.95cm} p{1.05cm}  p{0.85cm}  p{0.85cm} p{0.85cm} p{0.85cm}p{0.99cm} | p{0.75cm}p{0.75cm} }
        \hline
        & Score \rule{0pt}{2ex}& \small{BM25+T}\rule{0pt}{2ex} & \small{PS} \rule{0pt}{2ex}& \small{PF}\rule{0pt}{2ex}   & \small{Voyage}\rule{0pt}{2ex}  &\small{GritLM}\rule{0pt}{2ex}& \small{\tool(grit) }\rule{0pt}{2ex} & \small{SFR }\rule{0pt}{2ex}& \small{\tool(sfr) }\rule{0pt}{2ex} & t\rule{0pt}{2ex} & p\rule{0pt}{2ex}\\
       \cline{1-12}

        \multirow{9}{*}{\textbf{PatchFinder}} 
        & \textbf{MRR}        & 0.037&     0.010& 0.037&  0.433 & 0.452&     0.472& 0.458& \textbf{0.515}$\dagger$& -4.00&1e-4\\ \cline{2-12}
        & \textbf{Rec@10}    &0.062&    0.021& 0.059&  0.649
 & 0.679&   \textbf{0.744}$\dagger$&0.658 & 0.735$\dagger$& -3.70& 2e-4\\
        & \textbf{Rec@100}    &    0.261& 0.081& 0.219& 0.841
 & 0.820&   \textbf{0.847}&0.818 &
0.844& -1.86&0.85\\
        & \textbf{Rec@1000}  &    0.601&  0.356& 0.550& 0.917
 & 0.906 &   \textbf{0.916}&0.909 &
0.911& 0.53&0.59\\
        & \textbf{Rec@5000} &    0.929&  0.917& 0.925& 0.954
 & 0.952 &   \textbf{0.957}& 0.957 &
0.941& 1.89&0.059\\ \cline{2-12}
        & \textbf{NDCG@10}   &    0.019&  0.002& 0.018&  0.317
 & 0.331 &   0.312& 0.343 &
\textbf{0.387}& -2.95& 0.003\\
        & \textbf{NDCG@100}   &    0.020&  0.005& 0.023& 0.319
 & 0.332 &    0.313& 0.343 &
\textbf{0.388}& -2.96& 0.003\\
        & \textbf{NDCG@1000}  &    0.025& 0.024& 0.046& 0.330
 &0.340 &  0.321& 0.352 &
\textbf{0.395}& -2.97& 0.003\\
        & \textbf{NDCG@5000}  &    0.043& 0.088& 0.106& 0.355
 &0.369 & 0.349& 0.377& \textbf{0.421}& -3.00& 0.003\\
        \cline{1-12} 

        \multirow{9}{*}{\textbf{GitHubAD}}  
        & \textbf{MRR}       &    0.055&  0.010&  0.037&  0.291
 &0.232 &    \textbf{0.371}$\dagger$&0.247 & 0.367$\dagger$&  -4.52& 0 \\ \cline{2-12}
        & \textbf{Rec@10}   &    0.085&  0.011&  0.054&  0.446
 & 0.367&   \textbf{0.577}$\dagger$& 0.389 & 0.525$\dagger$&  -3.76& 2e-4\\
        & \textbf{Rec@100}   &
        0.406&  0.051& 0.185&  0.632
 & 0.565&  \textbf{0.773}$\dagger$& 0.605& 0.764$\dagger$& -7.57& 0\\
        & \textbf{Rec@1000} &     0.769&  0.289& 0.515&  0.816
 & 0.814&  0.884$\dagger$& 0.830& \textbf{0.906}$\dagger$&  -7.28& 0\\
        & \textbf{Rec@5000} &    0.933&  0.686&  0.933&  0.924
 &0.928 &   0.933& 0.940 & \textbf{0.940}& -2.81& 0.005\\ \cline{2-12}
        & \textbf{NDCG@10}   &     0.028& 0.004&   0.020&  0.194
 &0.155&   \textbf{0.255}$\dagger$& 0.161&0.250$\dagger$&  -3.24& 0.001\\
        & \textbf{NDCG@100}  &      0.029&  0.007&  0.024&  0.196
 &0.156&    \textbf{0.256}$\dagger$& 0.162 & 0.252$\dagger$&  -3.24& 0.001\\
        & \textbf{NDCG@1000}  &     0.039&  0.020&  0.050&  0.207
 &0.168&   \textbf{0.267}$\dagger$& 0.174& 0.262$\dagger$&  -3.24& 0.001\\
        & \textbf{NDCG@5000}  &    0.072&  0.068& 0.114&  0.245 & 0.209&  \textbf{0.301}$\dagger$& 0.214& 0.298$\dagger$& -3.25& 0.001\\
        \hline
    \end{tabular}
\end{table*}

\subsection{RQ1: \tool's Retrieval Scores}
\label{sec:rq1}


\noindent \textbf{Comparing \tool with Baselines}. Table~\ref{tab:main_result} compares \tool with existing methods (except VFCFinder). From \tool consistently outperforms PatchScout and PatchFinder by a large margin, and also surpasses \texttt{voyage-3}, the top-1 model on major code embedding benchmarks~\cite{li2024coir,husain2019codesearchnet}, on Recall@10 and MRR on both datasets. Paired $t$-tests show that the improvements over \texttt{voyage-3}
are statistically significant.
The performance gain is more pronounced on GitHubAD,
possibly because the commit messages in PatchFinder are already well-written (Appendix~\ref{app:data_stat} Table~\ref{tab:stat} shows the commit messages in PatchFinder are longer), leaving less room for improvement through better code modeling. The overall performance of CodeXEmbed (\texttt{Salesforce/SFR-Embedding-Mistral}, i.e., SFR) and GritLM are similar. 

\noindent \textbf{Understanding the Improvement over PatchFinder}. One question is how much of \tool's improvement over PatchFinder is attributed to the \tool framework itself. To answer this question, in Table~\ref{tab:main_result}, we report the scores using the model only, i.e., the columns GritLM and SFR are the scores by embedding the top-512 tokens for each commit. We further perform paired t-tests between the vanilla model and the SPFinder, and use $\dagger$ to denote that SPFinder is significantly better. We can observe that while the gap between vanilla LLM vs PatchFinder is large, the gain of SPFinder's MRR and Recall@10 over the vanilla LLM is also significant, especially on the GitHubAD dataset. This indicates that the effectiveness of \tool is not only attributed to the LLM embeddings but also to its framework design.

\noindent \textbf{Error Analysis of \tool}. We conduct an empirical study on the failing cases of \tool in both datasets. 
We randomly sample 40 CVEs whose patching commit is ranked below 150. Based on the availability of identifying words (e.g., function name/class name/variable name) in the CVE, we define 3 classes of errors and manually label cases into them: (1) \textbf{CVE description does not contain any identifying words and detailed explanation  (37.5\%)}. For example, CVE-2023-35152: "\emph{any logged in user can add dangerous content in their first name field and see it executed with programming rights. Leading to rights escalation}". These words describe the exploit rather than the code behavior, thus it is challenging to retrieve the patch based on the CVE. (2) \textbf{CVE description contains identifying words not in the commit (42.5\%)}: For example, the description of CVE-2021-41079~\cite{CVE-2021-41079} mentions the identifying word "\emph{NIO2}", but it is not in the commit (Figure~\ref{fig:ner}). (3) \textbf{the CVE description includes identifying keywords that directly correspond to changes in the diff (20\%)}: For example, CVE-2021-23390 cites the vulnerable \texttt{U.set()} and \texttt{U.get()} methods, this failure may be due to the common function names get/set.

\noindent \textbf{Strategies for Improving \tool}. Based on the study results, we propose three strategies for improving \tool: (1) enriching CVE descriptions with additional identifying cues from other advisory sources;
(2) strengthening path embedding via more effective in-repository search (Section~\ref{sec:NER}); 
and (3) fine-tuning the LLMs using CVE data can generally help mitigating these errors.

\noindent \textbf{Comparison with VFCFinder}. 
\tool consistently outperforms VFCFinder, and the overall trend is consistent with our full-repository evaluation. 
Detailed results and analysis are provided in Appendix~\ref{app:vfcfinder}.
We further find that exact version tags cover only about 60\% of true patching commits, 
which explains the recall limitation of version-based filtering approaches 
(Appendix~\ref{app:version}).



\begin{tcolorbox}[colback=white,colframe=black,left=3pt,right=3pt,top=3pt,bottom=3pt]
    \textbf{RQ1-Part1 (Scalability challenge)}: How accurate is \tool for handling full repositories?\\ [1.5ex]
    \textbf{Answer}: \tool achieves 0.736 and 0.577 Recall@10 on our two datasets, significantly outperforming existing work and the SOTA code embedding model. The gain is attributed to both the usage of LLMs and the design of \tool. 
    \end{tcolorbox}

\begin{table}
    \centering
    
    \caption{Time/financial cost of \tool and baseline components for processing 10,000 commits per CVE. 
    }
    \label{tab:efficiency}
   
    \begin{tabular}{p{3cm} p{2.5cm} p{1cm}}
        \hline
       Method & Step & Time\\ \hline

       \multirow{1}{*}{\tool} & Indexing & 30min \\ \hline
              \multirow{2}{*}{PatchScout} & Feature Extraction & 9min \\
        & Learning & 2.6min \\ \hline
        PatchFinder & Indexing & 9 mins\\ \hline
        Voyage & Indexing & \$1.8 \\ \hline
        VFCFinder & Feature Extraction & 50min \\ \hline
    \multirow{3}{*}{\parbox{3cm}{\textbf{\tool query\\(actual time cost)}}}
  & Querying GritLM & 14s \\
      & Querying GritLM file & 72s\\ 
       & Path embedding & \$0.021 \\  \hline

    \hline
    
    \end{tabular}
\end{table}

\vspace{-4pt}
\subsection{RQ1: \tool's Time cost}

We report the time and financial cost of the main components of \tool and the baselines in Table~\ref{tab:efficiency}. The time cost is measured by the time taken to process 10,000 commits of one CVE. The financial cost is measured by the cost of using the commercial API for processing 10,000 commits. We benchmark the runtime of \tool, PatchScout, and PatchFinder on the same L40S node for consistency (Section~\ref{sec:experiment_setup}).

\noindent \textbf{Analysis on Time Cost}. From Table~\ref{tab:efficiency}, we can see that \tool takes about 84s to query every 10,000 commits, and 40 mins to index every 10,000 commits when using an L40S GPU, this time cost is manageable for practical deployment to large repositories. 

Notice the indexing cost is \textbf{a one time effort} and can be done offline. Indexing commits and querying CVEs in SPFinder are separated. The following article \cite{querying_indexing} explains why the indexing is done offline, therefore the long indexing cost doesn't affect the actual performance of \tool, and the time for each CVE relies solely on the time cost of the querying stage. 

\begin{tcolorbox}[colback=white,colframe=black,left=3pt,right=3pt,top=3pt,bottom=3pt]
    \textbf{RQ1-Part2 (Scalability challenge)}: How fast is \tool for handling large repositories?\\ [1.5ex]
    \textbf{Answer}: \tool is efficient, it takes about 84s to query every 10,000 commits, and less than 30 mins to index (i.e., offline) every 10,000 commits, which is practical for large repositories. The financial cost of \tool is negligible. 
    \end{tcolorbox}

\subsection{RQ2: Analysis on Handling the Long Context}

Since \tool uses hierarchical embedding, one question is: is hierarchical embedding with a shorter context window a good way of handling long contexts? Should we instead trade the file-level embeddings for a larger context window? To answer this question, we compare the retrieval scores of GritLM and CodeXEmbed (SFR) on \texttt{apache/tomcat} (from GitHubAD) and \texttt{ImageMagick/ImageMagick6} (from PatchFinder) when embedding the top 512 to 4096 tokens. The results of \texttt{apache/tomcat} are shown in Appendix~\ref{app:context_length}. Both \texttt{apache/tomcat} and \texttt{ImageMagick/ImageMagick6} show a similar trend: the score does not improve significantly when extending the context window size. A similar degradation has been observed in another long text retrieval task~\cite{long_context_degrade}. Considering the time cost, we conclude that hierarchically embedding each file diff with a smaller context window is a practically effective way of handling the long commit. 

\noindent \textbf{Discussion on GritLM's Long Context Degradation}. The long context degradation is more obvious in GritLM. We find that the training data~\cite{grit_training_data} for GritLM's retrieval model lacks long code examples: while GritLM's training data includes coding tasks such as splash question to SQL, their lengths are shorter; in contrast, CodeXEmbed~\cite{liu2024codexembed} is trained on code data. For GritLM, the long context degradation problem can be potentially alleviated by fine-tuning using CVE data. 

\begin{tcolorbox}[colback=white,colframe=black,left=3pt,right=3pt,top=3pt,bottom=3pt]
    \textbf{RQ2 (Long context challenge)}: How effective is hierarchical embedding for handling long context for patch tracing? \\ [1.5ex]
    \textbf{Answer}: Since long context degradation is observed on multiple repositories, hierarchical embedding with a smaller window is practically effective to handle long context. 
    \end{tcolorbox}

\subsection{RQ3: Tracing Real-World Patches}

To evaluate \tool's ability to discover new patches, we apply \tool on CVEs in the GitHub Advisory database~\cite{githubAD}  with missing patch links. For each CVE, we manually verify the top-30 results and submit pull requests to add the patch link. One example is CVE-2024-22031~\cite{CVE-2024-22031}, whose patch was found and merged based solely on \tool, without relying on CVE mentions in commit messages. To date, we have submitted and merged patches for 35 CVEs:
26 ranked at top-1, 5 within top-30 (including 3 with multiple patches), and 2 not found within top-500. The complete list of merged CVEs and patches can be found in: \url{ https://github.com/AnonySE26/SPFinder/blob/master/githubAD_PRs.md}.

\begin{tcolorbox}[colback=white,colframe=black,left=3pt,right=3pt,top=3pt,bottom=3pt]
    \textbf{RQ3}: Can \tool identify real-world missing patch links?\\ [1.5ex]
    \textbf{Answer}: \tool can successfully trace real-world missing patch links. To date, \tool has merged 35 CVE patches into the GitHub Advisory database.
    \end{tcolorbox}

\subsection{RQ4: Feature Contribution}
\label{sec:eval_context_length}

\noindent \textbf{Ablation Summary}. We conduct an ablation study by removing one feature group at a time (Table~\ref{tab:feature_set}) and retraining the ranker with the same hyperparameters. Overall, all feature groups contribute positively. The embedding-based features contribute the most to the final ranking performance, followed by time features and path features. Full ablation plots and detailed results are provided in Appendix~\ref{app:ablation}.

\noindent \textbf{NER Quality}. We manually evaluate the NER module used for path embedding and observe high precision with reasonable recall; detailed numbers are reported in Appendix~\ref{app:ner}.


\begin{tcolorbox}[colback=white,colframe=black,left=3pt,right=3pt,top=3pt,bottom=3pt]
    \textbf{RQ4}: How does each feature in \tool contribute to the final ranking score?\\ [1.5ex]
    \textbf{Answer}: All feature groups help; embedding features contribute the most, followed by time and path features.
    \end{tcolorbox}

\noindent

\section{Discussions and Future Work}
\label{sec:future}

\noindent \textbf{Reducing the Mis-Matching between CVE and Commit}. Although \tool achieves a higher recall on the GitHubAD and PatchFinder datasets, there remain challenges in bridging the mismatch between the CVE description and the commit diff code. For example, how to bridge the gap between the vulnerability type (e.g., "\emph{this could lead to an infinite loop and denial-of-service}") with the diff code containing the infinite loop? One potential approach is to fine-tune the LLM for embedding using pairs of (CVE, vulnerable function) from existing vulnerability datasets~\cite{Fan2020BigVul,wang2021patchdb,zhou2021finding}. This approach enables the model to learn the associations between vulnerable code and the vulnerability types.





\section{Conclusion}

In this work, we introduce \tool, a scalable and effective patch retrieval system for security vulnerabilities that can scale to large-scale repositories. \tool addresses two limitations in existing work: (1) how to handle the long context of commits; (2) how to scale to the full repository by balancing the effectiveness and efficiency. \tool proposes a hierarchical embedding framework to address the long context challenge and a three-phase framework to address the scalability challenge. Our experiment shows that \tool outperforms existing work for patch tracing~\cite{li2024patchfinder,tan2021locating,dunlap2024vfcfinder} by a large margin; it further outperforms VoyageAI, a commercial embedding model with state-of-the-art performance, on the Recall@10 and MRR, by 18\% and 28\% on our two datasets. The time cost of \tool is manageable for practical deployment to large repositories. Using \tool, we have successfully identified and merged the patching links for 35 CVEs in the GitHub Advisory Database.



\clearpage
\clearpage
\bibliographystyle{ACM-Reference-Format}

\bibliography{ScalPatchRetrieval}


\clearpage
\appendix
\section{Appendix}
\subsection{Existing Baseline Methods}
\label{app:baseline}

\textbf{PatchScout}~\cite{tan2021locating} is among the first works to investigate the problem of vulnerability patch tracing. PatchScout proposes 22 features to model the similarity between the query CVE and the candidate commit (e.g., the number of shared words, function names) and a learning-to-rank framework; 

\noindent \textbf{PatchFinder}~\cite{li2024patchfinder} is a more recent work, it leverages a two-phase process for the retrieval, utilizing language models, e.g., the CodeReviewer~\cite{li2022codereviewer} model and the BertScore~\cite{zhang2019bertscore} framework, as well as fine-tuning;

\noindent \textbf{VFCFinder}~\cite{dunlap2024vfcfinder} improves upon existing work by directly addressing the scalability challenge by filtering the candidate commits using the fixed version number. It further provides an easy-to-use interface for users to trace the patches.

\subsection{CVE-Patch Mismatch }
\label{app:patch_mismatch}

\begin{figure}[h]
    \centering
    \begin{subfigure}
           \centering  
           \includegraphics[width=1.0\linewidth]{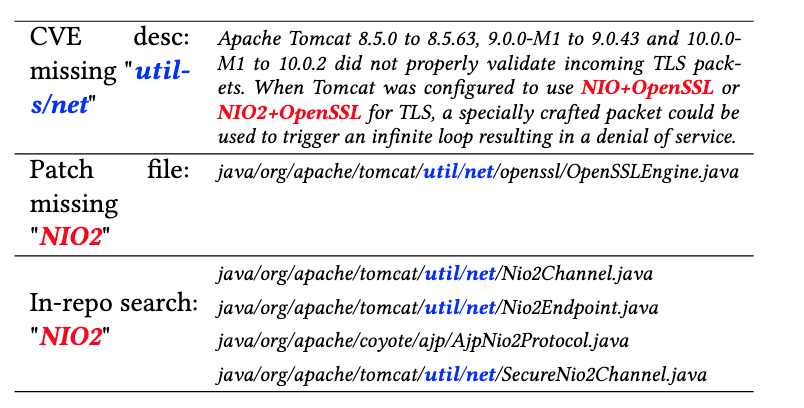}
           \includegraphics[width=1.0\linewidth]{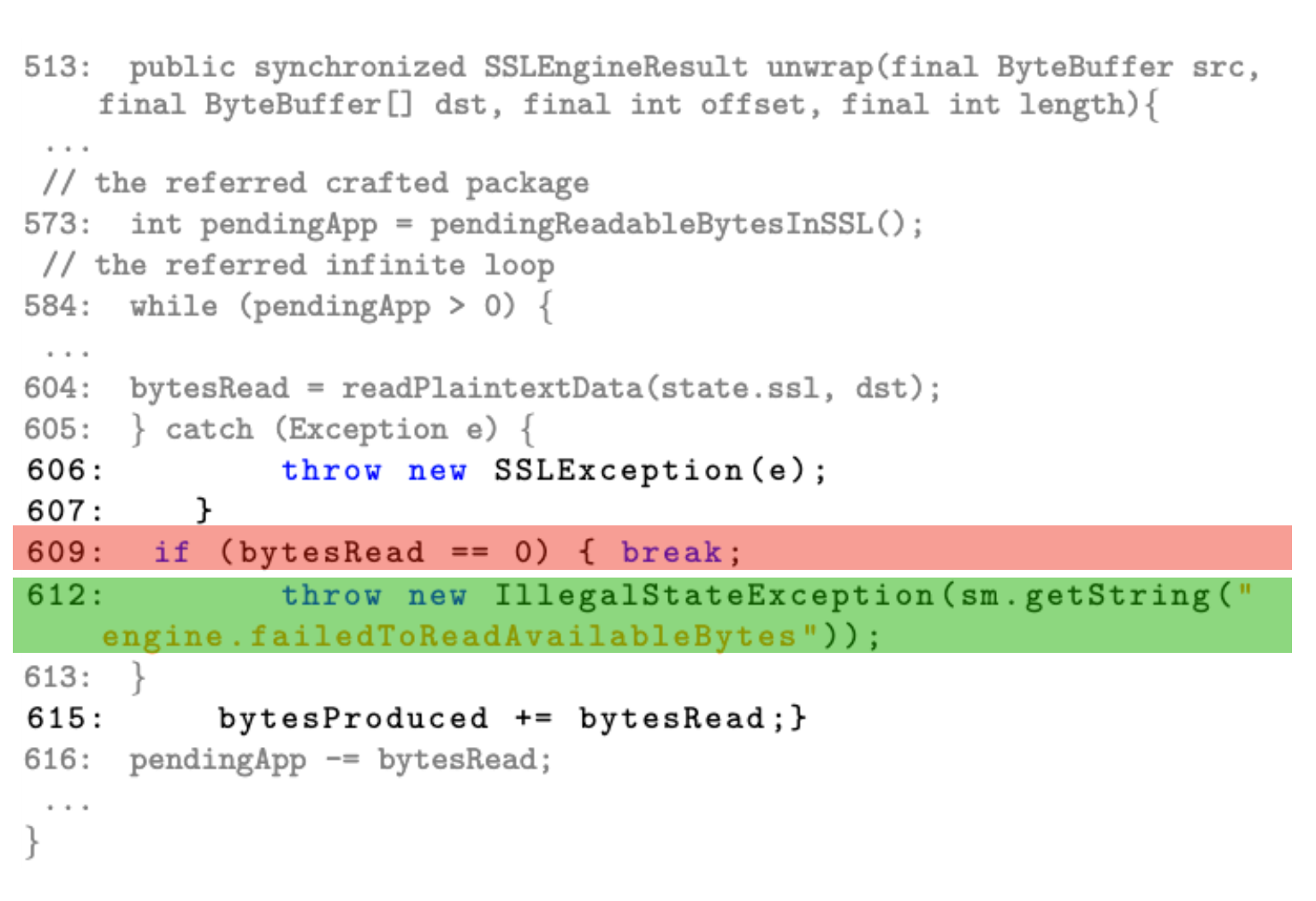}       \caption{An example of CVE-Patch Mismatch (CVE-2021-41079)~\cite{CVE-2021-41079}. The bottom listing shows the patching commit, where the grey texts are folded. As a result, there is no token overlap between the CVE description and the patch. \label{fig:ner} }
       \end{subfigure}
       
   \end{figure}

Figure~\ref{fig:ner} shows an example of CVE-Patch Mismatch.

\subsection{Dataset Statistics}
\label{app:data_stat}

Tables~\ref{tab:repo_cve_stats}--\ref{tab:stat} summarize the statistics of the
datasets used in this paper, including repository sizes, patch lengths,
token distributions, and train/test splits.

\begin{table}[h]
\centering
\caption{Statistics on the number of commits in each repo}
\label{tab:repo_cve_stats}
\begin{tabular}{lrrrr}
\toprule
\textbf{Metric} & \textbf{total} & \textbf{$>$5k} & \textbf{$>$10k} & \textbf{$>$5k portion}\\
\midrule
\#cves (GitHubAD)        & 6,789 & 3,316 & 2,420 & 49\%\\

\#cves (patchfinder)     & 4,795 & 4,029 & 3,845 & 84\%\\

\#repos (GitHubAD)       & 2,514 & 540  & 290 & 21\%\\
\#repos (patchfinder)    & 481   & 188  & 126 & 39\%\\
\bottomrule
\end{tabular}
\end{table}

\begin{table}[h]
\centering
\caption{Statistics on the number of tokens each patch contains. This table shows it is common for the patching commit to contain long context.}
\label{tab:patch_stats}
\begin{tabular}{lcc}
\toprule
\textbf{Metric} & \textbf{GitHubAD} & \textbf{patchfinder} \\
\midrule
\#patches            & 6.3k   & 4.8k   \\
avg \#token          & 24,021  & 1,286   \\
\% patches $>$ 512   & 71.8\% & 44.5\% \\
median \#token       & 1,059   & 441    \\
1/4 quartile         & 2,572   & 910    \\
3/4 quartile         & 450    & 254    \\
\bottomrule
\end{tabular}

\end{table}

\begin{table}[h]
    \caption{Comparison of token coverage \label{tab:context_size}}
    \centering
    \begin{tabular}{p{2.2cm}p{2.4cm}p{2cm} } \hline
    method & \#Tokens covered & Token coverage\\ \hline
   PatchFinder & 390 & 52\%\\ 
   VFCFinder & 390 & 52\%\\ 
   Voyage~\cite{voyage} & 2,600 & 82\%\\
   \tool & 1,820 & 67\% \\ \hline
    \end{tabular}
    \end{table}

\begin{table}[h]
    \caption{Statistics of the datasets used in this paper \label{tab:stat}}
    \centering
    \begin{tabular}{lll} \hline
    Statistics & PatchFinder & GitHubAD\\ \hline
    Tokens (commit msg)& 56 & 44\\ 
    Tokens per diff (white space) & 3,498 & 3,217\\
    Tokens per diff (voyage tokenizer) & 12,370 & 16,236\\
    Tokens per diff (GritLM tokenizer) & 14,293 & 19,241\\
    Tokens per file (GritLM tokenizer) & 2,454 & 2,577\\ \hline
    \# repos (train) & 343 & 477 (sampled) \\
    \# repos (test) & 101 & 148 \\
    \# CVEs (train) & 3,682 & 1,290 \\
    \# CVEs (test) & 629 & 952 \\ \hline
    Pos:neg & 1:20415 & 1:13449\\ \hline
    \end{tabular}
    \end{table}

\subsection{Model Selections for Embedding}
\label{app:model_select}

Table~\ref{tab:model_select} summarizes the detailed information for the embedding models used in this paper. The benchmark ranks refer to the CoIR leaderboard snapshot in Nov 2024.

\begin{table}[h]
    \caption{Selected LLMs for Embedding~\cite{muennighoff2022mteb} \label{tab:model_select}}
    \centering
    \begin{tabular}{p{3.1cm}p{1cm}p{0.5cm} p{0.5cm}p{1.9cm} } \hline
    Model & Context & Size & Rank & Fast inference?\\ \hline
   \texttt{voyage-3} & 32,000 & - & 1 & yes\\ 
   \texttt{gritLM/GritLM-7B} & unlimited & 7B & 2 & no\\ 
   \shortstack[l]{\texttt{Salesforce/}\\ \texttt{SFR-Embedding-Mistral}} & 32,768 & 7B & - & no\\ \hline
    \end{tabular}
    \end{table}

\subsection{Context Length vs. Score/Runtime}
\label{app:context_length}

\begin{figure}[h]
    \centering
    \begin{subfigure}
           \centering  
           \includegraphics[width=\linewidth]{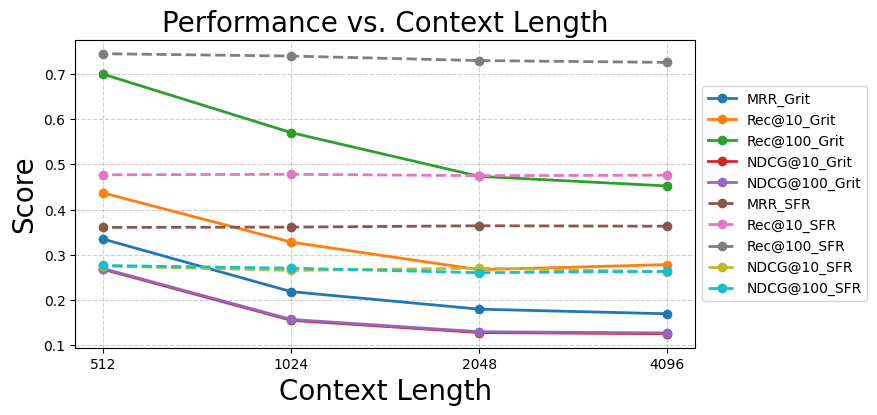}
           \includegraphics[width=\linewidth]{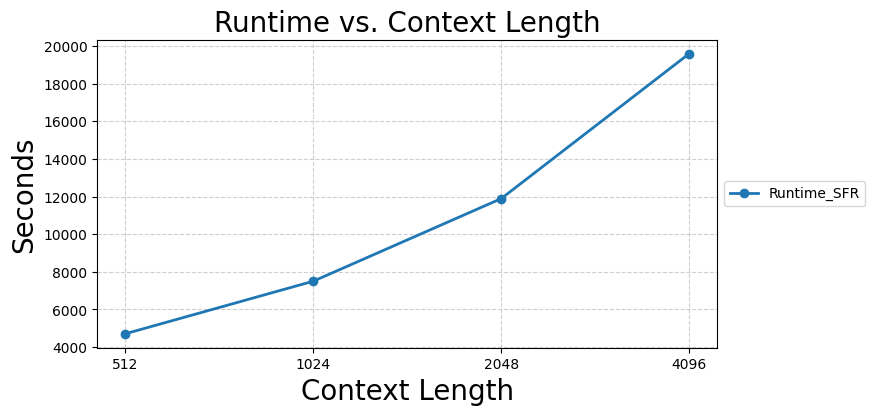}      
           
           \caption{Top: context length vs retrieval score, bottom: context length vs running time. We observe long context degradation~\cite{long_context_degrade}: the score is not significantly improved when using a longer context window (i.e., more information). \label{fig:context_length}}
       \end{subfigure}
       
   \end{figure}

Figure~\ref{fig:context_length} shows the effect of context length on retrieval
performance and runtime for GritLM and CodeXEmbed (SFR) on the
\texttt{apache/tomcat} dataset.

\subsection{Comparison with VFCFinder}
\label{app:vfcfinder}

\begin{figure}[h]
    \centering
    \includegraphics[width=0.9\linewidth]{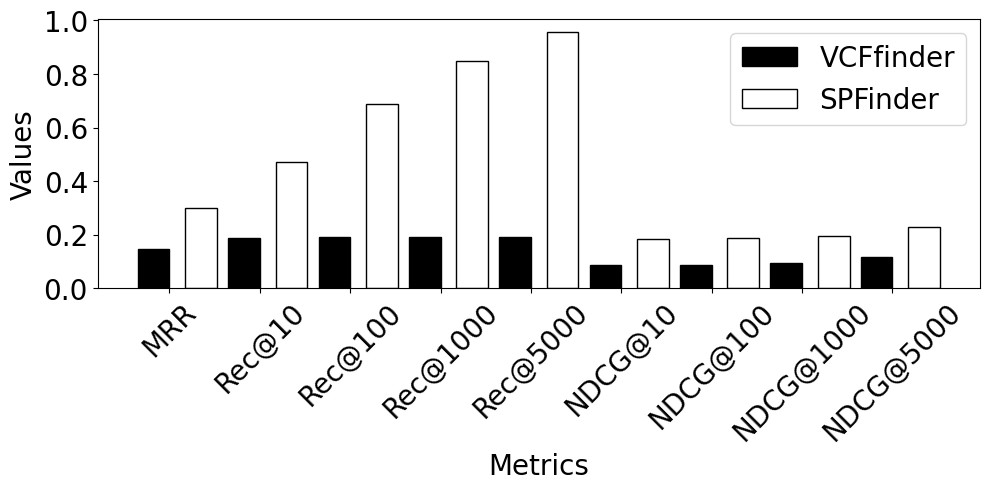}
    \caption{Comparison between \tool(grit) and VFCFinder~\cite{dunlap2024vfcfinder} on 8 repositories in the GitHubAD (234 CVEs) (Section~\ref{sec:baseline}) \label{fig:vfcfinder} }
\end{figure}

Figure~\ref{fig:vfcfinder} compares \tool and VFCFinder on a subset of
8 large repositories from GitHubAD (234 CVEs), which we use due to the
higher computational cost of VFCFinder (Table~\ref{tab:efficiency}). We can see that \tool outperforms VFCFinder on all metrics. In particular, VFCFinder's recall stays the same across all top k. This is because VFCFinder only retrieves a small set of commits within the fixed version, thus the recall will not keep increasing. By inspecting the output of VFCFinder, we find that many CVEs (60\%) are not processed successfully due to the failure in its pipeline to handle the fixed version or the fixed version is not available. 

\subsection{Correctness of Version-Based Filtering}
\label{app:version}

Table~\ref{tab:distance_recall} summarizes an empirical analysis of the
correctness of version tags in CVE descriptions.
We analyze 2,614 NVD CVEs with known patches and measure how far the
mentioned fix version must be expanded to include the true patching commit.
Exact version tags cover only 60.24\% of patches, and recall improves
gradually as the version window expands.
This result indicates that version-based filtering is prone to low recall
when used alone for patch tracing.

\begin{table}[h]
    \centering
    \caption{Statistics of the version distance between the mentioned fix version and the true fix version  \label{tab:distance_recall}}
    \begin{tabular}{p{1.2cm}p{1.2cm}p{1.5cm}} \hline
    Range & Count & Recall (\%) \\ \hline
    0 & 1,677& 60.24 \\
    $\pm$1 & 1,875 & 67.35 \\
    $\pm$5 & 2,136 & 76.72 \\
    $\pm$50 & 2,614  & 93.89 \\ \hline
\end{tabular}
\end{table}

\subsection{Ablation Study Details}
\label{app:ablation}

This subsection reports detailed ablation results for \tool(grit).
We remove one feature group at a time from the full model and retrain
using the same hyperparameters (selected by FLAML), following the feature
group definitions in Table~\ref{tab:feature_set}.

\begin{figure}
    \centering
    \begin{subfigure}
        \centering    
\includegraphics[width=0.9\linewidth]{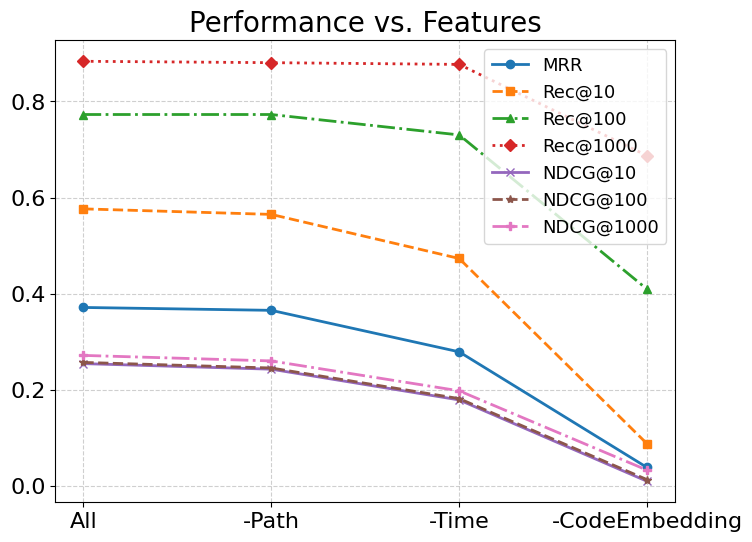}\\
    \end{subfigure}
    \begin{subfigure}
        \centering    
\includegraphics[width=0.9\linewidth]{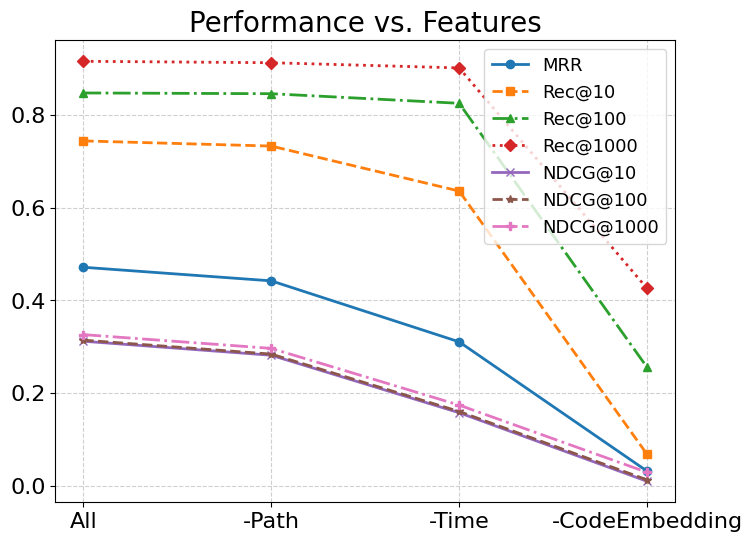}
    \end{subfigure}

    \caption{\tool(grit)'s performance changes (MRR, Recall@K, and NDCG@K) on the GitHubAD (top) and PatchFinder dataset (bottom), when different feature groups are incrementally removed. 
     \label{fig:ablation_study} }
\end{figure}

Figure~\ref{fig:ablation_study} shows the effect of incrementally removing
different feature groups on GitHubAD and PatchFinder.
All feature groups contribute positively to performance.
In particular, removing embedding features leads to the largest drop in recall,
followed by time features and path features, indicating the importance of both
hierarchical embedding and path embedding in \tool(grit).

\begin{figure}
    \centering
    \begin{subfigure}
        \centering    
\includegraphics[width=1.0\linewidth]{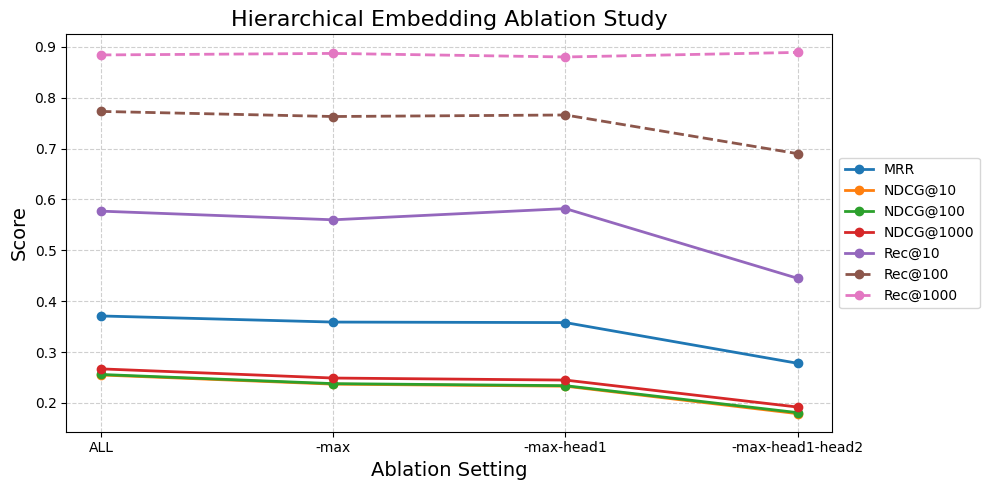}
    \end{subfigure}

        \begin{subfigure}
        \centering    
\includegraphics[width=1.0\linewidth]{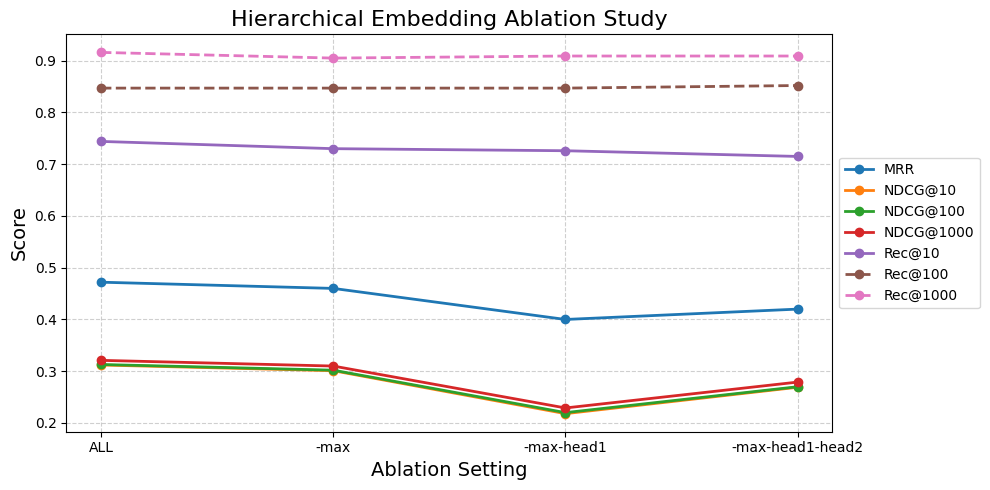}
    \end{subfigure}
    
    \caption{\tool(grit)'s ablation study on the specific feature in the hierarchical embedding group: GitHubAD (top) and PatchFinder dataset (bottom), when different feature groups are incrementally removed. 
     \label{fig:ablation_feature} }
\end{figure}

Figure~\ref{fig:ablation_feature} further presents feature-level ablation
results within the hierarchical embedding group for both datasets,
providing a more fine-grained analysis of individual feature contributions. 

\subsection{NER Accuracy for Path Embedding}
\label{app:ner}

Table~\ref{tab:evaluate_ner} reports the accuracy of the NER component used for path embedding (Section~\ref{sec:NER}).
Precision indicates that a CVE description contains no incorrectly annotated
entities, while recall indicates that all correctly annotated entities are
included in the GPT-generated results.
Overall, the NER annotations achieve perfect precision and reasonably high
recall across both datasets.

\begin{table}[h]
    \centering
    \caption{Manual annotation results of NER accuracy\label{tab:evaluate_ner} }
    \begin{tabular}{p{2.2cm}p{1.2cm}p{1.5cm}} \hline
     &  Precision & Recall \\ \hline
     AD train & 1.0 & 0.81 \\ 
     AD test & 1.0 & 0.82 \\ 
     PatchFinder train & 1.0  &  0.85\\ 
     PatchFinder test & 1.0  &  0.76\\ \hline
\end{tabular}
\end{table}

\subsection{Hyperparameters}
\label{app:hyperparameters}

\noindent \textbf{Phase-1 Pre-ranking (Scalability).}
In Phase~1 (Section~\ref{sec:3_phase}), we incorporate temporal information
from the MITRE CVE database~\cite{cve}, including publish time and reserve time.
The scores of commit message BM25, diff code BM25, reserve time, and publish time
are combined using a weighted sum.
The weights are selected via grid search on the training data, with the final
weights set to [0.35, 0.15, 0.30, 0.20], respectively.

The grid search space is:
commit message BM25 $\in \{0.30, 0.35, 0.40\}$,
diff code BM25 $\in \{0.10, 0.15, 0.20\}$,
reserve time $\in \{0.20, 0.30, 0.40\}$,
and publish time $\in \{0.20, 0.30, 0.40\}$.
To ensure score comparability, each component score is converted to its
reciprocal rank (i.e., $1/\text{rank}$) before aggregation.

\noindent \textbf{Phase-3 Learning-to-Rank.}
In Phase~3, we run FLAML for approximately one hour (about 30 trials) to
automatically identify the optimal hyperparameters.
For the GitHubAD dataset, the selected parameters are:
learning rate = 0.01, num\_leaves = 30, and min\_data\_in\_leaf = 38.
For the PatchFinder dataset, the selected parameters are:
learning rate = 0.01, num\_leaves = 15, and min\_data\_in\_leaf = 10.

\subsection{Why We Do Not Compare with PatchFinder Phase-2}
\label{app:patchfinder_phase2}

PatchFinder Phase-2 re-ranks only the top-100 candidates returned by Phase-1 (Section~\ref{sec:baseline} in~\cite{li2024patchfinder}).
Therefore, Phase-2 recall is upper-bounded by Phase-1 Recall@100.

Quantitatively, Table~\ref{tab:main_result} shows that PatchFinder Phase-1 achieves Recall@100 $<0.219$ on our two datasets.
Thus, more than $1-0.219=78.1\%$ of true patching commits are discarded before Phase-2, making the Phase-2 recall necessarily $<0.219$ under full-repository evaluation.

A concrete example is CVE-2021-41079 (\texttt{apache/tomcat}), whose patch is commit \texttt{34115fb3c83f6cd97772232316a492a4cc5729e0}.
PatchFinder ranks the true patch at position 2485, while \tool ranks it at position 375.
We provide the full outputs and reproduction instructions at:
\url{https://github.com/AnonySE26/SPFinder/tree/master/example-CVE-2021-41079}
and a step-by-step guide to reproduce PatchFinder at:
\url{https://github.com/AnonySE26/SPFinder/tree/master/baselines},
based on their official implementation:
\url{https://github.com/MarkLee131/PatchFinder/tree/main/PatchFinder/TF-IDF}.


\subsection{Illustrative Example for Hierarchical Feature Construction}
\label{app:feature_values}

An illustrative example is shown in Table~\ref{tab:feature_values}.
F2 is the maximum cosine similarity among the top-5 BM25-ranked files.
F3 uses the top-1 file embedding.
F4 uses the mean embedding of the top-2 files.

\begin{table}[h]
\centering
\caption{Example file-level scores used to construct hierarchical features.
}
\label{tab:feature_values}
\setlength{\tabcolsep}{4pt}
\begin{tabular}{lccccc}
\toprule
 & \textbf{BM25} & \textbf{GritLM} & \textbf{F2} & \textbf{F3} & \textbf{F4} \\
\midrule
File 1 & \textbf{1.0}  & 0.3 & --  & 0.3 & 0.35 \\
File 2 & \textbf{0.5}  & \textbf{0.4} & 0.4 & --  & --   \\
File 3 & 0.3  & 0.2 & -- & -- & -- \\
File 4 & 0.2  & 0.3 & -- & -- & -- \\
File 5 & 0.1  & 0.2 & -- & -- & -- \\
File 6 & 0.01 & 0.6 & -- & -- & -- \\
\bottomrule
\end{tabular}
\end{table}

\end{document}